\documentclass[particles,article,submit,pdftex,moreauthors]{Definitions/mdpi} 

\firstpage{1} 
\pubvolume{1}
\issuenum{1}
\articlenumber{0}
\pubyear{2024}
\copyrightyear{2024}
\datereceived{ } 
\daterevised{ } 
\dateaccepted{ } 
\datepublished{ } 
\hreflink{https://doi.org/} 

\Title{Neuromorphic Readout for Hadron Calorimeters}

\TitleCitation{Title}

\Author{Enrico Lupi $^{1,2}$ \orcidI{}, Abhishek$^{7}$\orcidB{}, Max Aehle$^{6,11}$\orcidJ{}, Muhammad Awais $^{1,2,3,11}$\orcidE{}, Alessandro Breccia$^2$ \orcidZ{}, Riccardo Carroccio $^{2}$\orcidR{}, Long Chen$^{6,11}$\orcidL{}, Abhijit Das$^{9}$\orcidC{}, Andrea De Vita$^{1,2}$\orcidD{}, Tommaso Dorigo $^{1,3,4,11}$\orcidT{}, Nicolas R. Gauger$^{6,11}$\orcidK{}, 
Ralf Keidel$^{8,11}$\orcidW{}, 
Jan Kieseler $^{8}$\orcidF{}, Anders Mikkelsen$^{9}$\orcidM{}, Federico Nardi$^{2,10}$ \orcidN{}, Xuan Tung Nguyen$^{1,6}$\orcidH{}, Fredrik Sandin $^{3,11}$\orcidS{}, Kylian Schmidt$^{8}$\orcidG{}, Pietro Vischia$^{4,5,11}$ \orcidA{} 
, Joseph Willmore$^{1}$\orcidX{} 
}  

\AuthorNames{Enrico Lupi, Abhishek, Max Aehle, Muhammad Awais, Alessandro Breccia, Riccardo Carroccio, Long Chen, Abhijit Das, Andrea De Vita, Tommaso Dorigo, Nicolas R. Gauger, 
Ralf Keidel, 
Jan Kieseler, Anders Mikkelsen, Federico Nardi, Xuan Tung Nguyen, Fredrik Sandin, Kylian Schmidt, Pietro Vischia, 
Joseph WIllmore
}

\AuthorCitation{Lupi, E.; Abhishek; Aehle, M.; Awais, M.; Breccia, A.; Carroccio, R.; Chen, L.; De Vita, A.; Dorigo, T.; Gauger, N.R.; 
Keidel, R.; 
Kieseler, J.; Mikkelsen, A.; Nardi, F.; Nguyen, X.T.; Sandin, F.; Schmidt, K.; Vischia, P. Willmore J.
}

\address{$^{1}$\quad INFN, sezione di Padova - Via F. Marzolo 8, 35131 Padova, Italy \\
$^{2}$ \quad Universit\`a di Padova, dipartimento di Fisica e Astronomia, Via F. Marzolo 8, 35131 Padova, Italy \\
$^{3}$ \quad Lule{\aa} University of Technology, 971 87 Lule\aa, Sweden \\
$^{4}$ \quad Universal Scientific Education and Research Network, Italy \\
$^{5}$ \quad Universidad de Oviedo and ICTEA, Spain \\
$^{6}$ \quad University of Kaiserslautern-Landau (RPTU), Gottlieb-Daimler-Straße, 67663 Kaiserslautern, Germany \\
$^{7}$ \quad National Institute of Science Education and Research, Jatni, 752050, India\\
$^{8}$ \quad Karlsruhe Institute of Technology, 76131 Karlsruhe, Germany \\
$^{9}$ \quad Department of Physics and NanoLund, Lund University, Lund, Sweden \\
$^{10}$ \quad Laboratoire de Physique Clermont Auvergne, 63170 Aubière, France \\
$^{11}$ \quad MODE Collaboration, \url{https://mode-collaboration.github.io}\\
$^*$ \quad Corresponding author: \texttt{enrico.lupi@studenti.unipd.it} \\
}

\abstract{ 
We simulate hadrons impinging on a homogeneous lead-tungstate ($\text{PbWO}_4$) calorimeter to investigate how the resulting light yield and its temporal structure, as detected by an array of light-sensitive sensors, can be processed by a neuromorphic computing system. Our model encodes temporal photon distributions as spike trains and employs a fully connected spiking neural network to estimate the total deposited energy, as well as the position and spatial distribution of the light emissions within the sensitive material. The extracted primitives offer valuable topological information about the shower development in the material, achieved without requiring a segmentation of the active medium. A potential nanophotonic implementation using III-V semiconductor nanowires is discussed. It can be both fast and energy efficient.}

\begin{document}


\section {Introduction \label{s:intro}}

Hadron calorimeters play a critical role in high-energy physics experiments, providing precise energy measurements of hadronic showers.
A hadron calorimeter is a block of dense matter that is capable of stopping highly energetic particles through the strong interaction they withstand in hitting nuclear matter, and produce a macroscopic effect that suitable electronics can record; the latter is usually constituted by the release of visible light in an amount proportional to the incident particles energy. Over the past sixty years, the technology of hadron calorimeters evolved significantly. Still, the emphasis on the design of these instruments has stayed with the reduction of uncertainties in measuring the total incident energy of streams of hadrons. The reason for this ultimately lies within the strong interaction dynamics, which turn energetic quarks or gluons produced in high-energy processes (a scattering reaction or a decay of a heavy particle) into a collimated stream of hadrons, whose individual properties are less important to the experiment than their collective ones. 

While efforts to improve the energy measurement of hadronic jets through the individual measurement of its components date back to the 1980iess~\cite{zeusef}, it was only proven after the turn of the century how the combination of a precise tracking of charged hadrons and the association with different energy deposits in finely segmented calorimeters could enormously improve the overall energy measurement of the jet~\cite{Sirunyan_2017}. At the same time, it was realized how the hadronic decays of massive particles such as $W$ and $Z$ bosons, top quarks, and $H$ bosons could be distinguished from the otherwise insurmountable Quantum Chromodynamics (QCD) backgrounds if the originating particles were highly boosted. The discrimination again required a fine segmentation of the calorimeter and the exploitation of subtle topological features of the energy distribution within the cone of the resulting wide jets~\cite{kasieczka2018}.

A high segmentation of the sensitive elements of a calorimeter constitutes a significant challenge due to its impact on total cost, energy consumption, and output data volume. In particular, conventional computing solutions alone may not be sufficient to cope with the computing demand of online dimensional reduction of the resulting data in the foreseeable future.
Another significant challenge lies in the temporal realm.
Since particles traveling close to the speed of light (c) traverse 3 cm in just 100 ps (picoseconds), the time frame for pattern detection is condensed to sub-nanoseconds. Current detectors have not exploited such temporal information, which could help extract information about particle identity and help the overall event reconstruction.
Integrating an efficient online dimensional reduction and a spatio-temporal pattern recognition using parallel analog-digital neuromorphic computing architectures integrated in the detector volume could assist in overcoming these limitations. \\

Neuromorphic technologies and compute-in-memory architectures are ideal for processing streaming sensor data, especially when combined with event-based detectors and asynchronous distributed algorithms~\cite{Kudithipudi2025}. This is because they use the physical properties of materials and circuits to process information in parallel with high energy and latency performance~\cite{mehonic2024neuromorphic}. Furthermore, neuromorphic architectures encode information in physical time, such as through a succession of binary events called spikes, which offer additional modalities for encoding information besides bits, see for example~\cite{Nilsson2023a,Nilsson2023b}.
Spiking Neural Networks (SNNs) are used to model such hybrid analog-digital neuromorphic systems, and biological neurons using differential equations describing the dynamics. Unlike conventional artificial neural networks (ANNs), which use continuous values to represent the activation of the individual computational units/neurons, the computational units in SNNs generate discrete spikes in response to incoming stimuli. These spikes occur asynchronously at precise points, adding the temporal dimension to neural processing that enhances the model's capacity to asynchronously process time-dependent information efficiently. This spike-based approach makes SNNs particularly well-suited for event-based spatiotemporal processing, where spikes' dynamics and asynchronous parallel processing of spikes are efficiently implemented using specialized algorithms, circuits, devices, and materials.
A variety of mathematical models, such as the Leaky Integrate-and-Fire (LIF) model and the (adaptive) exponential integrate-and-fire model, are used to model the computational units in SNNs depending on the model capacity and level of biological realism required. These models vary in complexity but share the core concept that information encoded in the timing of spikes is processed via the responses of nonlinear analog dynamical systems.

This work is structured as follows. In Sec.~\ref{sec:data_gen}, we describe the data generation process, including the simulation setup, the calorimeter model, and the assumptions made for light production and readout. Section~\ref{sec:taks} defines the primary regression tasks and the target variables used for event characterization. In Sec.~\ref{sec:snn}, we introduce the neuromorphic computing approach, detailing the preprocessing steps, SNN architecture, and training protocol. Section~\ref{sec:results} presents the results of the regression models, including both single- and multi-target performance evaluations, along with an optimization study of the network architecture and hyperparameters. Section~\ref{sec:hardware_Implementation} explores a potential nanophotonic hardware implementation using III-V semiconductor nanowires, discussing its feasibility for high-speed and energy-efficient neuromorphic processing. Finally, in Sec.~\ref{sec:conclusion}, we summarize our findings, outline the advantages of neuromorphic readout for hadron calorimetry, and discuss potential future directions, including experimental validation and further hardware developments.

\section {Data generation}
\label{sec:data_gen}

To carry out this study, we use \texttt{GEANT4}~\cite{GEANT4} simulations of the response of a homogeneous calorimeter hit by single 100-GeV charged hadrons (either $p$, $K^+$, or $\pi^+$). The data were initially produced for the work described in Ref.~\cite{pidcalo}. The same data are used for this study after they are further processed to simulate light production, propagation, and detection (see {\em infra}).

\subsection {Calorimeter model}

The homogeneous calorimeter is made of Lead Tungstate ($\text{PbWO}_4$), which has a light yield equal to $LY_{PWO} = 200 ph/MeV$~\cite{light_yield}. It has a total size of $300 \times 300 \times 1200 \, \text{mm}^3$, which corresponds to a lateral width of $7.66 \rho_M$ (Molière radii) and $5.92 \lambda_I$ (interaction lengths), ensuring an average lateral containment of 100\% and a longitudinal containment of approximately 87\%. The detector
can be logically divided into a grid of $10\times10\times10$ units called cubelets, with dimensions of $3 \times 3 \times 12 \, \text{cm}^3$. Each cubelet is itself organized into a $10\times10\times10$ grid of cells, with dimensions of $3 \times 3 \times 12 \, \text{mm}^3$, for a total of one million cells in the whole calorimeter. \\

\subsection {Readout model} 

The readout system is made up of a system of light-sensitive sensors, one per each cubelet. They are placed on their upper face, on the $xz$ plane, and organized in a grid of $10\times10$. Each sensor has the same size of a cell in the $x$ and $z$ directions, while its height in the $y$ direction is considered negligible. The sensors are assumed to be 100\% efficient, so that they always detect all incident photons, regardless of their frequency. This simplification abstracts away from the hardware problem of manufacturing an effective readout system, in order to allow us to focus on the problem of information processing. Possible hardware implementations of such a system are discussed in Sec.~\ref{sec:hardware_Implementation}.


\subsection {Simulation of light production}

In order to simulate light production and diffusion inside the calorimeter, we make the following assumptions:
\begin{enumerate}
    \item As detailed in Ref.~\cite{pidcalo}, the position of each interaction is recorded only using the index of the cell where it occurred. All interactions are treated as if they occurred in the center of their respective cell.
    \item All deposited energy is converted into photons with 100\% efficiency.
    \item No Cherenkov radiation is emitted.
    \item Photons are emitted uniformly in all directions.
    \item The detector material is considered completely transparent to the photons, and they travel unimpeded until they reach the face of the cubelet where they were produced. They are absorbed by the cubelet faces, and the hit is registered only if the photons reach the upper one, where the photosensitive sensors are placed. 
\end{enumerate}
Concerning the above assumptions, in particular the 100\% efficiency of light collection, it is important to point out that the goal of this study is not to model all detector physics in detail, but rather to evaluate whether useful topological information can be extracted from the temporal structure of light emission. The assumption of perfect light yield removes hardware-dependent inefficiencies and allows for a clear focus on algorithmic performance. In a real experimental setup, the efficiency of light production is usually determined through calibration measurements, and systematic corrections can be applied to account for realistic light yield inefficiencies. Thus, this assumption does not invalidate the feasibility of the approach. Any systematic bias introduced by this assumption would be absorbed in the learned parameters of the SNN. If needed, the network can be retrained later with realistic efficiency factors.

Given the above assumptions, we simulate the light production as follows. For each interaction, we convert the energy deposited into the number of photons produced: $N_{ph}^{tot} = E_{dep} \cdot LY_{PWO}$. Given the cell in which the interaction occurred, we then calculate the solid angle subtended by each light sensor of the cubelet from the center of the cell, $\Omega^{i}$, and the time it takes for the light to reach it, $\Delta t^{i}$. The amount of photons that reach the $i$-th sensor after an interaction is thus simply given by $N_{ph}^{i} = N_{ph}^{tot} \frac{\Omega^{i}}{4\pi}$, while the arrival time is given by $t^{i} = t_{interaction} + \Delta t^{i}$. 

\subsection {Discretization of light pulses}

The arrival of the photons is recorded in a time window spanning from the start of the event up to $20 \, \text{ns}$: by this time, all major interactions have already occurred and more than 90\% of the total energy has been deposited in the calorimeter. Time is discretized into 100 bins of $0.2 \, \text{ns}$ each, and the total number of photons received within each of them is recorded. 

\subsection{Primary vertex}

For this study, we chose not to include all cubelets in our analysis, but to focus only on the one where the first nuclear interaction vertex occurred, which typically contains a large share of the released energy, and offers the highest information content on the identity of the incident particle\cite{pidcalo}. \\

The primary particles were originally produced in a fixed position, at a distance of $3 \,\text{m}$ from the calorimeter surface along the $z$ direction, and at its center on the $xy$ plane. This meant that the position of the first nuclear interaction vertex showed very little variation between events, which translated into little variation in the centroid of the energy deposition, one of the variables to be regressed (see Sec.~\ref{sec:taks} for the exact definition). 

To have a more realistic setting with a randomly distributed energy centroid, we introduce $\textit{a posteriori}$, a random shift in the initial position of the particle gun. This is practically done by adding to the position of each interaction random shifts along $x$ and $y$, both generated uniformly in the range $\left[0; \, 3 \right]  \, \text{cm}$. This shift is kept fixed for all interactions in the same event.

\subsection {Data samples}

For each simulated particle ($p,\pi^+,k^+$), a total of 5,000 simulated events were analyzed, recording the number of photons detected by each light sensor at each time step in the primary vertex cubelet. \\

\section{Task definition}
\label{sec:taks}

In this work we focus on regressing the following variables of interest: the total energy deposited in the calorimeter cubelet, the spatial coordinates of the centroid of the energy depositions,  and their relative dispersions. Each of these sets of variables will be regressed independently in the so-called single-target regression tasks, or together with other variables in multi-target regression. \\

The first variable of interest is the total amount of energy released inside the cubelet, to demonstrate that a neuromorphic system would be able to mimic the results obtained by traditional calorimeters.

The actual values of the total energy released span several orders of magnitude, such that directly regressing them would be a too challenging task. To circumvent the problem at this stage, we apply a non-linear transformation and select as objective of the regression the logarithm in base 10 of the energy expressed in MeV.

\begin{equation}
    log(E/MeV) = \log_{10}\left(\sum_iE_{dep}^i [MeV] \right)
\end{equation}

\begin{figure}[H]
    \centering
    \includegraphics[width=0.6\linewidth]{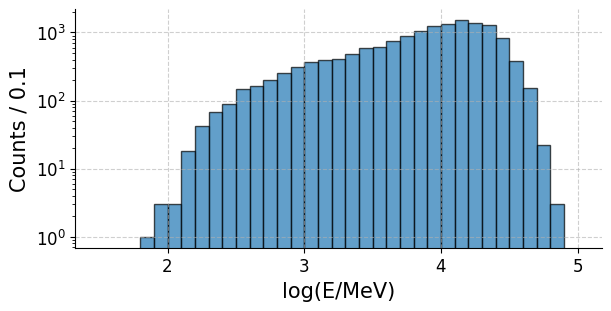}
    \caption{Distribution of the $log\text(E/MeV)$ variable across all the dataset.
    \label{fig:logE_distr}}
\end{figure}

The other variables are instead related to the geometry of the event. Their exact definition is the following:
\begin{equation}
    X_c = \cfrac{\sum_i X_i \cdot E_i}{E_{tot}},
\end{equation}
\begin{equation}
    \sigma^2 = \cfrac{\sum_i \left(X_i - X_c \right)^2 \cdot E_i}{E_{tot}},
\end{equation}
where the index $i$ runs over each interaction happening inside the cubelet, and the two variables are computed separately for each coordinate $\vec{X} = \left\{ x, y, z\right\}$.

It is important to note that both variables are expressed in units of "cell lengths": the $X_i$ variables range from 0 to 9 depending on the cell of the cubelet where the interaction occurred. \\

\begin{figure}[H]
    \centering
    \includegraphics[width=0.9\linewidth]{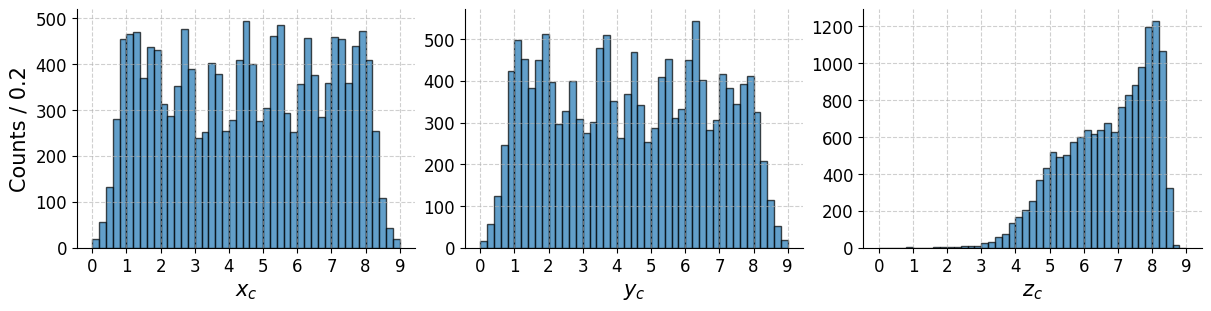}
    \\[1em] 
    \includegraphics[width=0.9\linewidth]{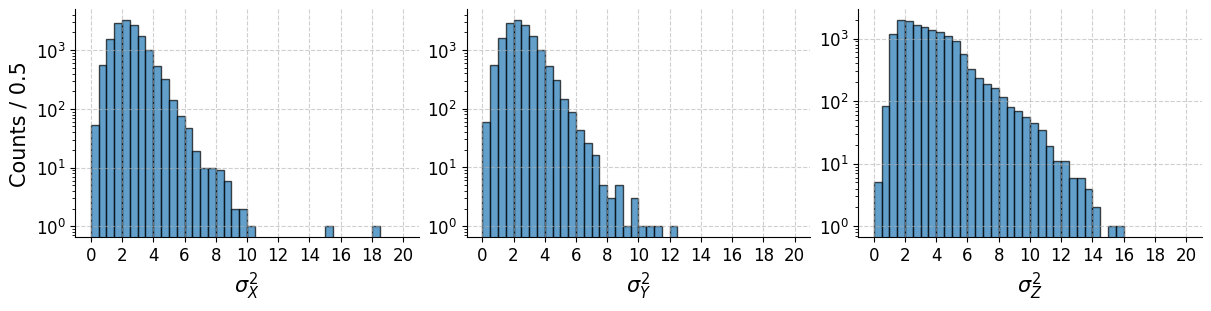}
    
    \caption{Distributions of the energy deposition centroid (above) and energy dispersion (below) across all the dataset.}
    \label{fig:centr_sigma_distr}
\end{figure}

\section {Spiking Neural Network model}
\label{sec:snn}

\subsection{Preprocessing: encoding procedure}

Incoming data need to be converted into a series of spikes, i.e. a spiketrain, before being fed into the network.
We adopt the following encoding scheme. For each sensor, we consider four different channels that go out of it and into the network. In each time step, these channels can either be activated and carry a spike or be inactive, depending on the amount of light received in that instant. Each channel has a different activation threshold, so that spikes will carry information about both timing and intensity of the interactions. The exact formula used is the following:
\begin{equation}    
S^{(i)}[t] = 
\begin{cases} 
       1, & \text{if  $N_{ph}[t] \geq 10^{i+2}$} \\
       0, & \text{otherwise} \\
\end{cases}
,\;\; i = 0, 1,2,3
\label{eqn:spike}
\end{equation}
where the index $i$ refers to the channel, while $t$ indicates the timestep.

\begin{figure}[H]
\centering
\includegraphics[width=8 cm]{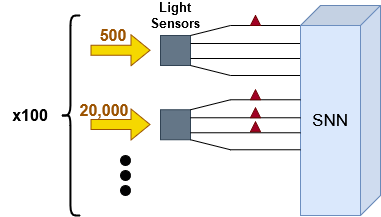}
\caption{Graphical representation of the encoding scheme.\label{fig1}}
\end{figure}

\subsection {Network architecture}
We use a fully connected feed-forward network with 400 input afferents, two hidden layers of 120 neurons each, and an output layer of varying total size depending on the task, with populations of 20 neurons per target. We adopt a simple LIF model as the neuron model, using learnable $U_{thr}$ and $\beta$ parameters, and $arctan$ function as surrogate gradient. The neuromorphic elements of the code were implemented using the $snnTorch$ Python library. ~\cite{Eshraghian2023}

The reason why this particular architecture was adopted is discussed in Sec.~\ref{sec:optimization}.

\subsection {Decoding schemes}

Neuromorphic systems offer many solutions to perform regression tasks, using either the output spikes or the membrane potential of the last layer as quantities to be analyzed. The latter is a continuous value, thus offering higher precision at the cost of higher latency and computational complexity. The former, on the other hand, needs more refined decoding schemes to handle its binary nature (rate-based, latency-based, spiking interval codes), but usually provides better performances. \\

A possible decoding scheme using the membrane potential value consists of simply equating its value at the last time step to the variable to regress. In order to make this scheme more stable, we can consider a population of output neurons and take the average of their membrane potential:
\begin{equation}
    \hat{X} = \cfrac{1}{N_{pop}}\sum_{i=1}^{N_{pop}}U^{(i)}[T]
\end{equation}

In contrast, a promising scheme that uses the output spikes is the rate-based encoding, which links the variable to regress to the number of spikes produced.
Using once again a population of neurons to stabilize the results, we obtain the following expression:
\begin{equation}
    \hat{X} = \cfrac{1}{N_{pop}}\sum_{i=1}^{N_{pop}}\sum_{t=1}^{T}S^{(i)}[t]
\end{equation}

In this paper, we will show the results obtained using the second decoding scheme outlined.

\begin{figure}[H]
\centering
\includegraphics[width=13.5cm]{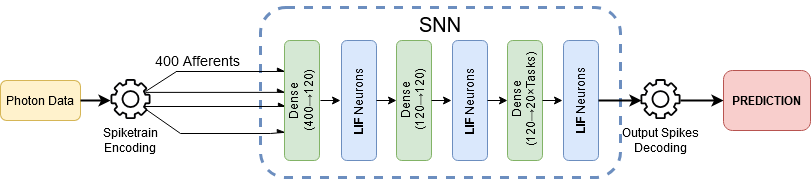}
\caption{Data processing pipeline and network architecture.
\label{fig:net_schema}}
\end{figure}

\subsection{Training}

The complete dataset is divided into training, validation and test datasets with a 0.7-0.15-0.15 split. The validation dataset was used for hyperparameter optimization (Sec.~\ref{sec:optimization}) and to monitor the loss behavior during training. \\

Different loss functions were adopted depending on the distributions of the target variable. For $log(E/MeV)$, the centroid $z$ coordinate and the energy dispersion, we use the Mean Absolute Error (MAE) loss (or L1 loss), given the small dynamical range of the outputs and their asymmetrical distribution, as shown in Figs.~\ref{fig:logE_distr}--~\ref{fig:centr_sigma_distr}. For the energy centroid $x$ and $y$ coordinates, instead, we adopt the Mean Squared Error (MSE) loss. This is because optimizing the MSE loss corresponds to calculating the batch distribution's mean, while for the L1 loss it corresponds to the median of the batch distribution, which is a better estimate for a skewed distribution. \\
In case of multi-target regression tasks, we compute the losses for each target individually and then sum them to obtain the final loss estimate. \\

We used Adam~\cite{Adam} as optimizer, with a learning rate $\eta$ and a weight regularization $\lambda$ of $0.001$, and parameters $\beta_1=0.9, \, \beta_2=0.999$.
Together with it, we used a learning rate scheduler, that decreases $\eta$ each epoch of a factor $\gamma=0.7$. \\

Networks were trained for 5 epochs using a batch size of 50 samples.

\section {Results}
\label{sec:results}

In this section, we will cover the results obtained by the neuromorphic system both on single and multi-target regression, and the procedure followed to optimize the architecture hyperparameters.

When reporting the error, we use the average of the $l_1$ distance between prediction and target across all the test dataset, eventually weighted by the target itself.

\begin{equation}
    \epsilon_{rel} = \cfrac{1}{N_{test}} \sum_i \left|\frac{target_i - prediction_i}{target_i} \right| 
\end{equation}
\begin{equation}
    \epsilon_{abs} = \cfrac{1}{N_{test}} \sum_i \left| target_i - prediction_i \right| 
\end{equation}

\subsection{Single-target regression}
\label{sec:single-target}

\begin{figure}[H]
    \centering
    \includegraphics[width=0.4\linewidth]{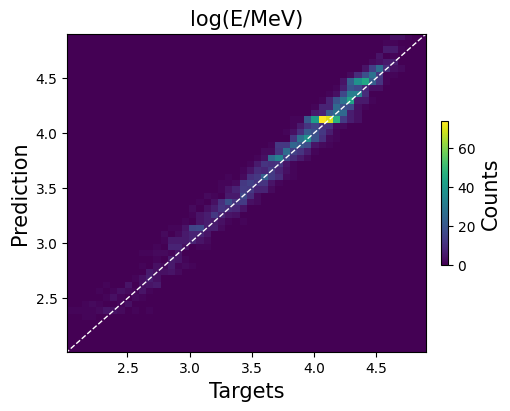}
    \includegraphics[width=0.4\linewidth]{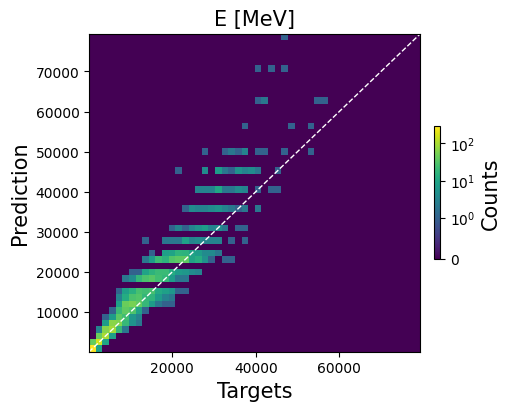}
     \\[1em]
    \includegraphics[width=0.4\linewidth]{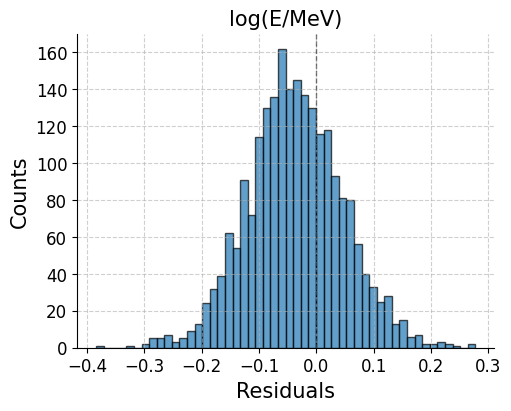}
    \includegraphics[width=0.4\linewidth]{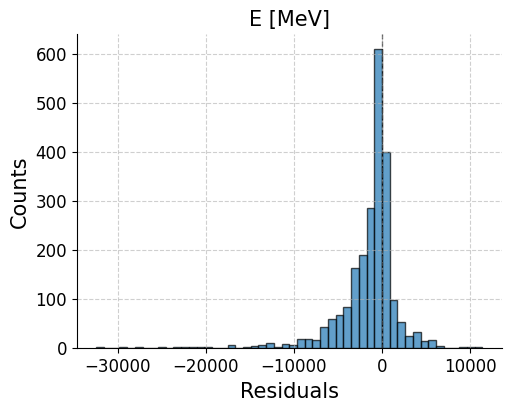}
    
    \caption{Output of the $log(E/MeV)$ network, and the respective results for the true value of the energy. The plots show the correlation between targets and predictions (above) and the residuals (below).}
    \label{fig:reuslts_E}
\end{figure}

\begin{figure}[H]
    \centering
    \includegraphics[width=0.32\linewidth]{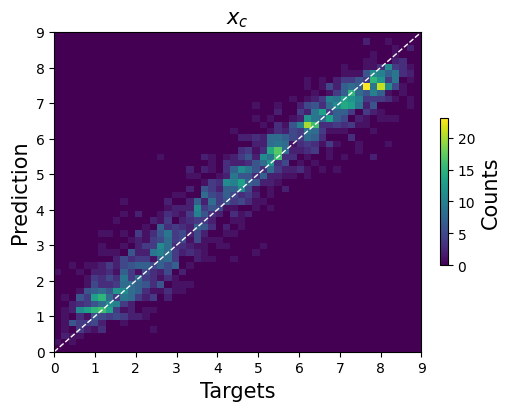}
    \includegraphics[width=0.32\linewidth]{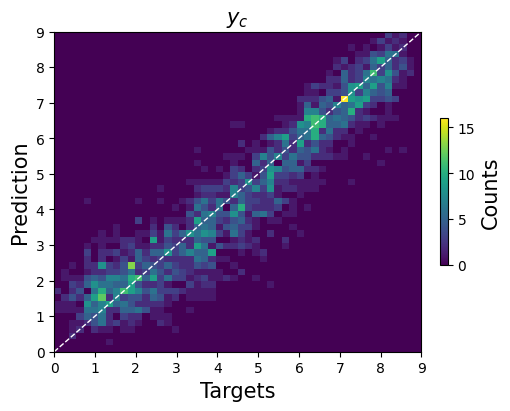}
    \includegraphics[width=0.32\linewidth]{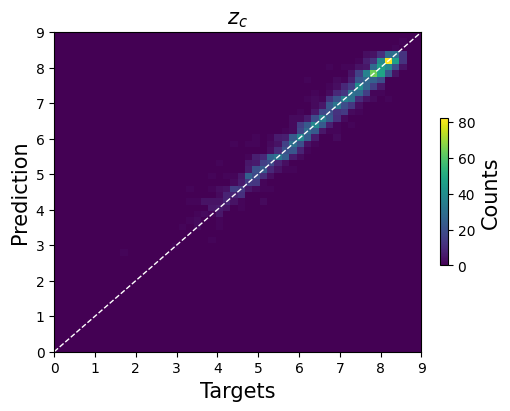}
    \\[1em]
    \includegraphics[width=0.32\linewidth]{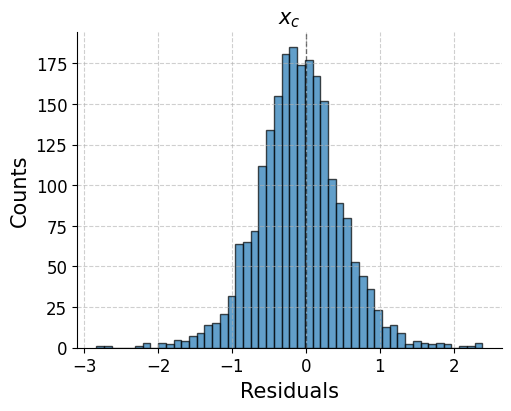}
    \includegraphics[width=0.32\linewidth]{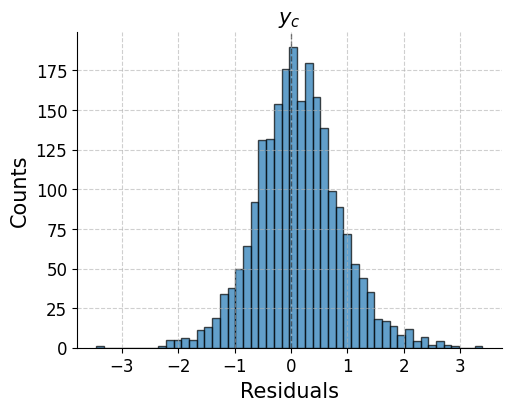}
    \includegraphics[width=0.32\linewidth]{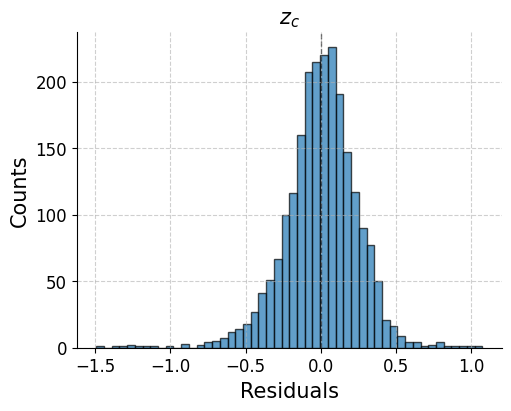}
    
    \caption{Output of the energy centroid network, for the $x$, $y$ and $z$ coordinates. The plots show the correlation between targets and predictions, (above) and the residuals (below).}
    \label{fig:reuslts_centr}
\end{figure}

\begin{figure}[H]
    \centering
    \includegraphics[width=0.32\linewidth]{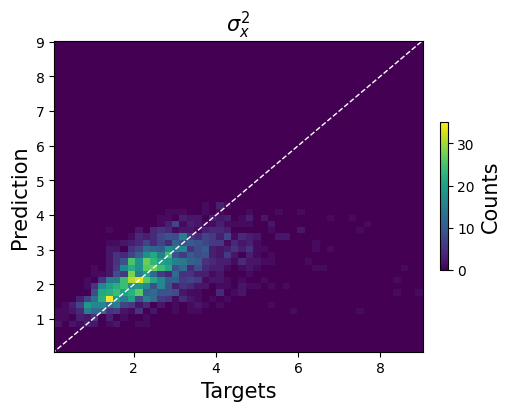}
    \includegraphics[width=0.32\linewidth]{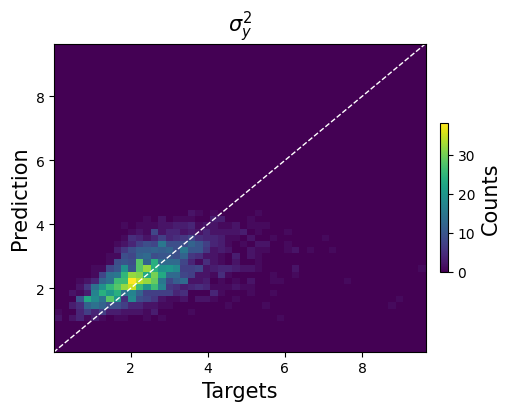}
    \includegraphics[width=0.32\linewidth]{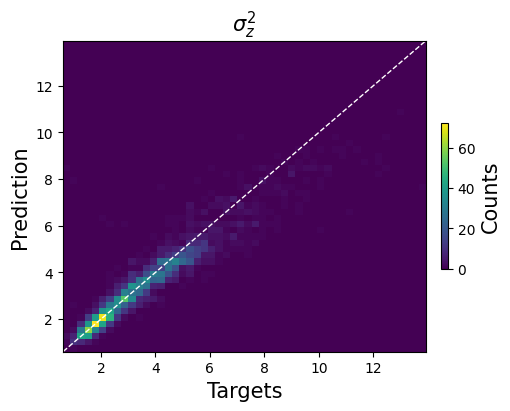}
    \\[1em]
    \includegraphics[width=0.32\linewidth]{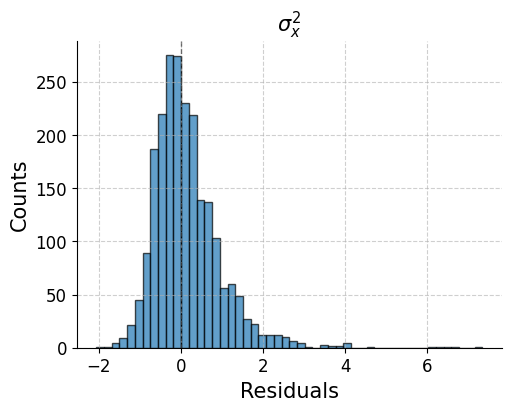}
    \includegraphics[width=0.32\linewidth]{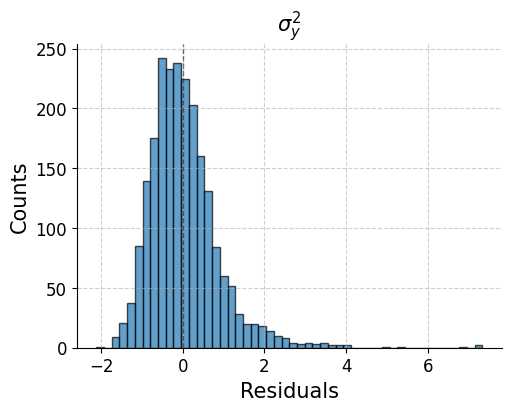}
    \includegraphics[width=0.32\linewidth]{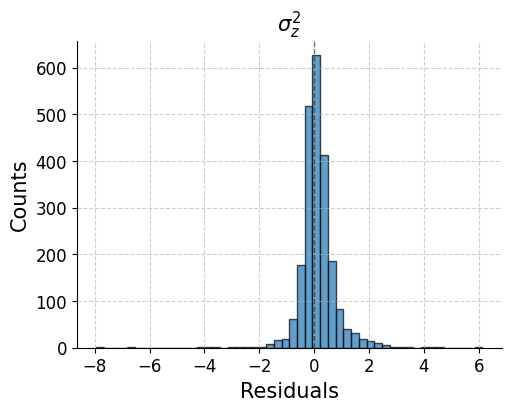}
    
    \caption{Output of the energy dispersion network, for the $x$, $y$ and $z$ coordinates. The plots show the correlation between targets and predictions (above) and the residuals (below).}
    \label{fig:reuslts_sigma}
\end{figure}

Table ~\ref{tab:single-target} shows the results obtained for single-target regression tasks, while Figs.~\ref{fig:reuslts_E}--~\ref{fig:reuslts_sigma} show the comparison between targets and predicted values, and the respective residuals. Given the large dynamical range of the energy deposition in the cubelets we considered (see Fig.~\ref{fig:logE_distr}, and the regression to the logarithm of the energy, the absolute error on the regressed energy is large, as it is an average over the whole sample. We also observe that the position of the centroid of the energy distribution can be estimated by the SNN with good accuracy (with an uncertainty of less than one cell in all coordinates). For the dispersions of the energy distribution in the cubelets, we obtain good results for the $z$ component, while the $x$ and $y$ dispersions are less easy to evaluate. 

\begin{table}[H] 
\caption{Single-target regression errors. Position and variance estimates are expressed in cell unit lengths. \label{tab:single-target}}
\newcolumntype{C}{>{\centering\arraybackslash}X}
\begin{tabularx}{\textwidth}{CCCCCCCCC}
\toprule

	& \textbf{log(E/MeV)}
    & \textbf{E}
    & \textbf{$x_c$} 
    & \textbf{$y_c$} 
    & \textbf{$z_c$} 
    & \textbf{$\sigma^2_x$} 
    & \textbf{$\sigma^2_y$} 
    & \textbf{$\sigma^2_z$} \\
\midrule
    $\epsilon_{rel} (\%)$
    & 1.975
    & 18.37
    & 18.13
    & 24.74
    & 2.85
    & 26.07 
    & 31.39 
    & 12.04 \\
    $\epsilon_{abs}$
    & 0.073
    & 2039 MeV
    & 0.44
    & 0.58
    & 0.18
    & 0.58 
    & 0.62 
    & 0.41 \\
\bottomrule
\end{tabularx}
\end{table}

\subsection {Multi-target regression}
\label{sec:multi-target}

\begin{figure}[H]
    \centering
    \includegraphics[width=0.49\linewidth]{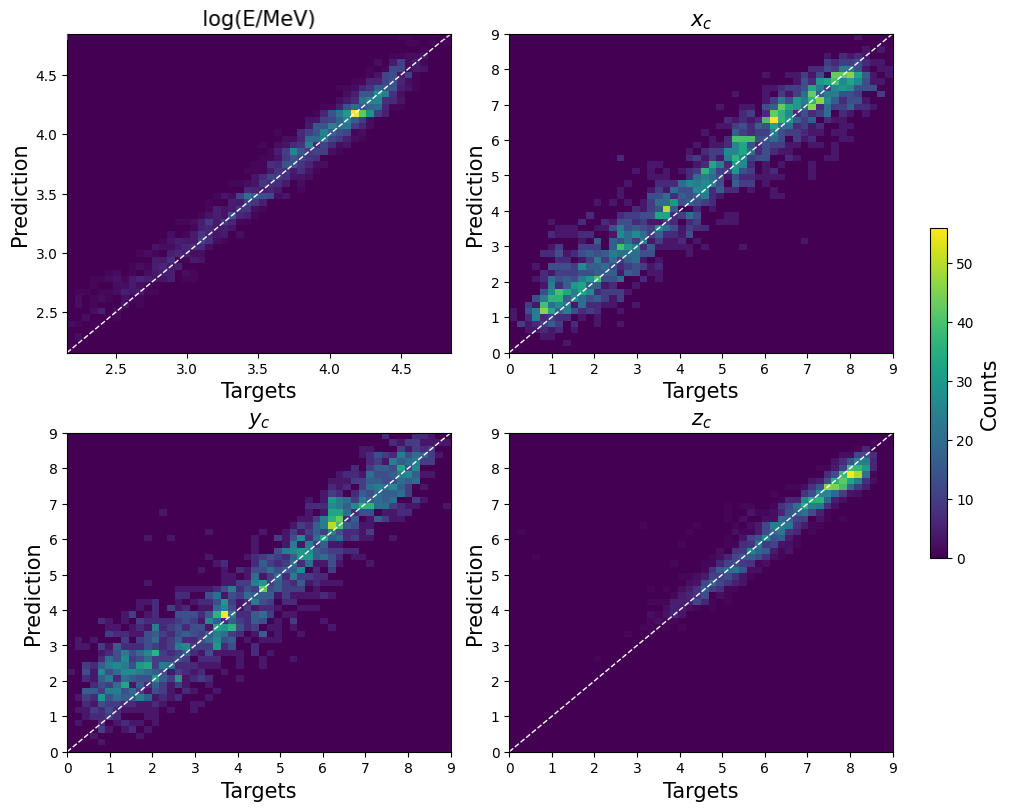}
    \includegraphics[width=0.49\linewidth]{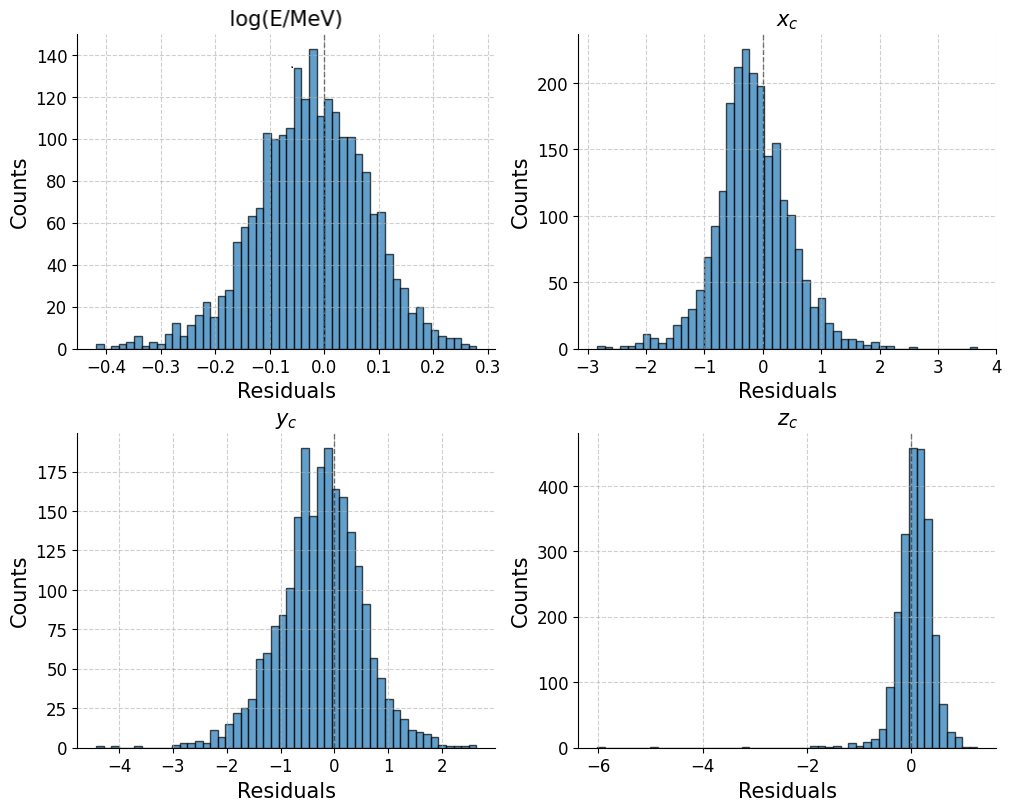}
    \\[1em]
    \includegraphics[width=0.3\linewidth]{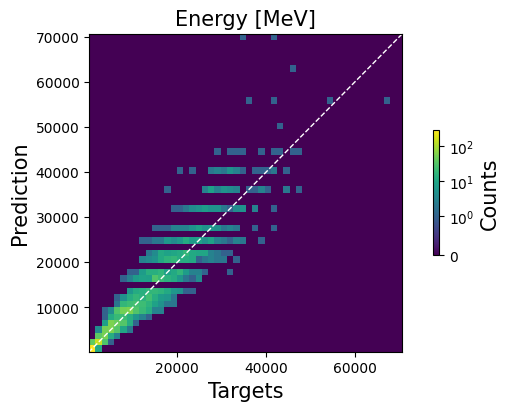}
    \includegraphics[width=0.3\linewidth]{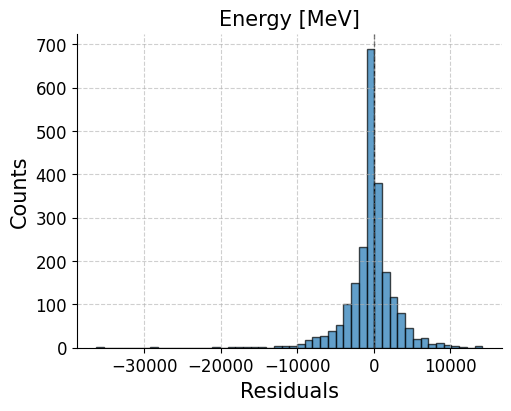}
    
    \caption{Output of the energy deposition and centroid network. The plots show the correlation between targets and predictions (left) and the residuals (right).}
    \label{fig:reuslts_E_pos}
\end{figure}

\begin{figure}[H]
    \centering
    \includegraphics[width=0.49\linewidth]{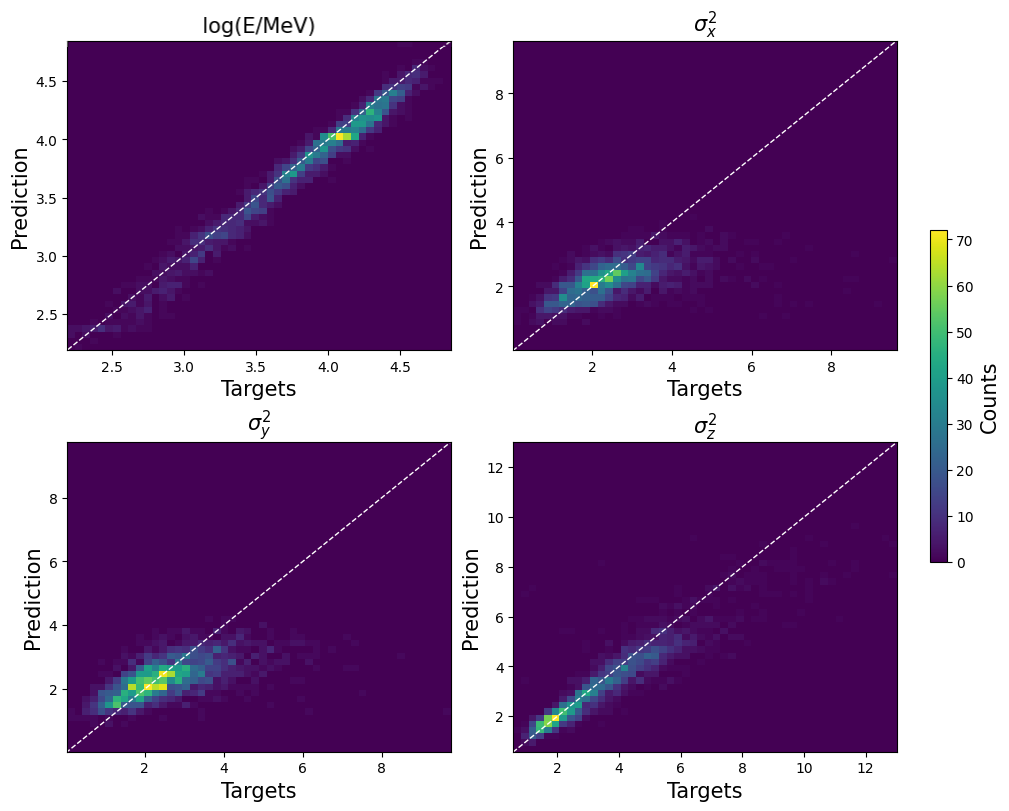}
    \includegraphics[width=0.49\linewidth]{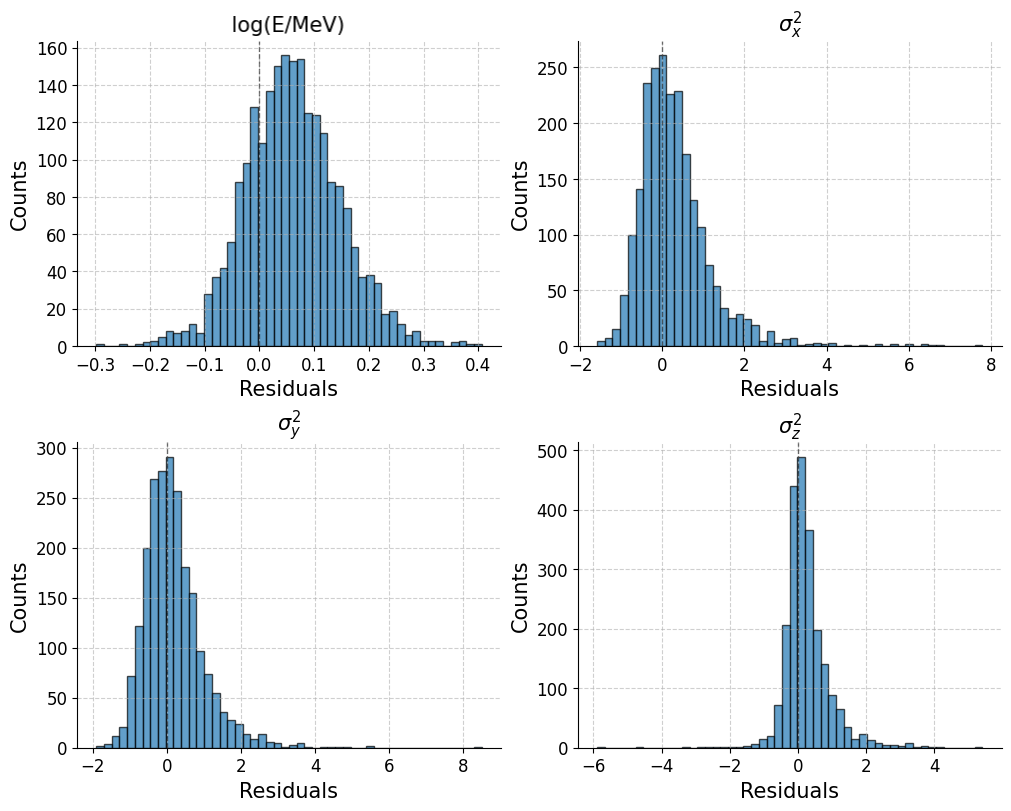}
    \\[1em]
    \includegraphics[width=0.3\linewidth]{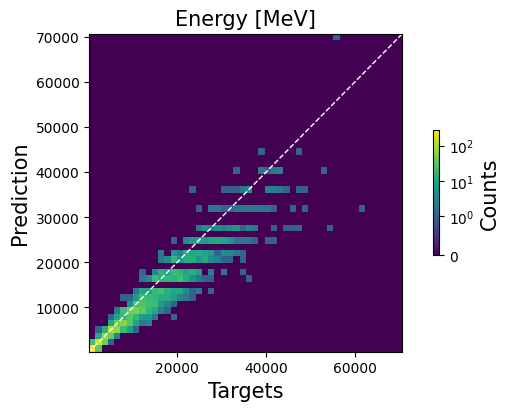}
    \includegraphics[width=0.3\linewidth]{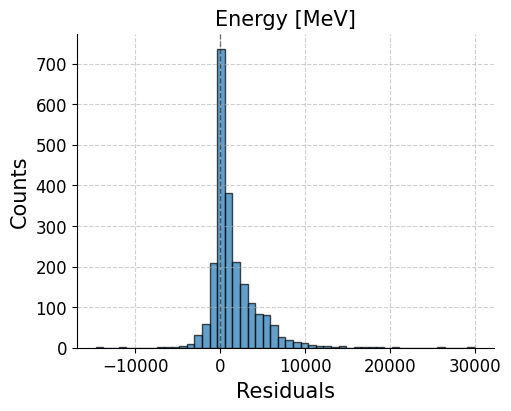}
    
    \caption{Output of the energy deposition and dispersion network. The plots show the correlation between targets and predictions (left) and the residuals (right).}
    \label{fig:reuslts_E_sigma}
\end{figure}

We trained two different networks: one regressing $log(E/MeV)$ and the position of the centroid, the other regressing $log(E/MeV)$ and the energy dispersion in all three directions. Table~\ref{tab:Epos_result} and Fig.~\ref{fig:reuslts_E_pos} show the results for the former network, while Table~\ref{tab:Esigma_results} and Fig.~\ref{fig:reuslts_E_sigma} concern the latter.

\begin{table}[H] 
\caption{Multi-target regression errors, for deposited energy and energy centroid.\label{tab:Epos_result}}
\newcolumntype{C}{>{\centering\arraybackslash}X}
\begin{tabularx}{\textwidth}{CCCCCCC}
\toprule

	& \textbf{log(E/MeV)}
    & \textbf{E}
    & \textbf{$x_c$}
    & \textbf{$y_c$}
    & \textbf{$z_c$} \\
\midrule
    $\epsilon_{rel} (\%)$
    & 2.262
    & 20.59
    & 20.66
    & 32.66
    & 5.33 \\
    $\epsilon_{abs}$
    & 0.082
    & 2001 MeV
    & 0.51
    & 0.62
    & 0.25 \\
\bottomrule
\end{tabularx}
\end{table}

\begin{table}[H] 
\caption{Multi-target regression errors, for deposited energy and energy dispersion.\label{tab:Esigma_results}}
\newcolumntype{C}{>{\centering\arraybackslash}X}
\begin{tabularx}{\textwidth}{CCCCCCC}
\toprule

	& \textbf{log(E/MeV)}
    & \textbf{E}
    & \textbf{$\sigma^2_x$} 
    & \textbf{$\sigma^2_y$} 
    & \textbf{$\sigma^2_z$} \\
\midrule
    $\epsilon_{rel} (\%)$
    & 2.252
    & 17.21
    & 27.71
    & 26.40
    & 12.76 \\
    $\epsilon_{abs}$
    & 0.084
    & 1967 MeV
    & 0.63
    & 0.59
    & 0.47 \\
\bottomrule
\end{tabularx}
\end{table}

The results of multi-target regressions are consistent with those shown {\it supra} from single-target regression, indicating that the used model is capable of handling both tasks with similar performance.


\subsection {Optimization of architecture and hyperparameters}
\label{sec:optimization}

To estimate the best network architecture, a Bayesian optimization algorithm was implemented and applied to both the multi-target regressions involving energy together with centroids and the one involving energy and dispersions. 
The scheme is based on a Gaussian surrogate model of the objective ``black box''objective function. This model acts as prior on the function space, which gets updated by new observations. The latter are generated to maximize a so-called acquisition function, that in our case is the Expected Improvement. This method directs the sampling towards the maximum of the surrogate Gaussian model, allowing us to obtain a faster optimization, due to conditioning on sampling towards maxima, and gives also a posterior distribution of the performance over the parameters, which is useful to obtain further insights on the total parameters space considered. Since the optimization aims to sample towards maxima, we defined a measure of performance that quantify how well a certain architecture or parameters are doing. Performance P is defined as the mean of relative error reciprocals:
\begin{equation}
    P = \frac{1}{N_{targets}} \sum_i^{N_{targets}}\frac{1}{\epsilon_i}
\end{equation}
where $N_{targets}$ is the total number of targets we are regressing together and $\epsilon_i$ is the relative error on target "i" regression. The errors are computed on the validation set.

\begin{figure}[H]
    \centering
    \includegraphics[width=0.4\linewidth, height = 5cm]{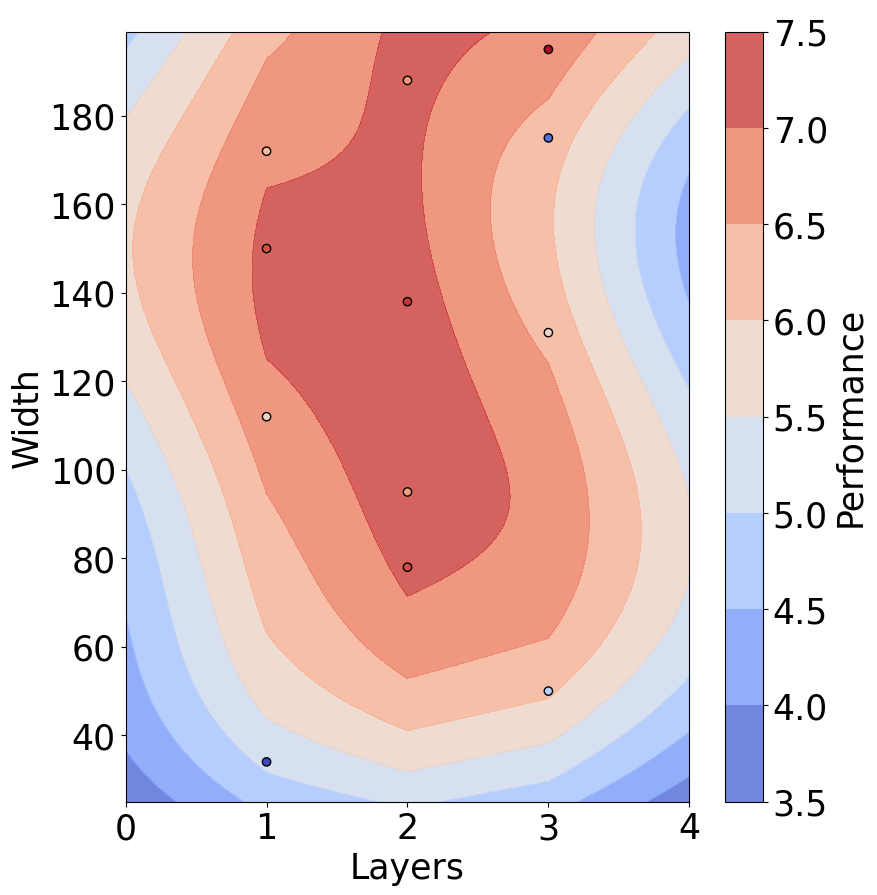}
    \hspace{1cm}
    \includegraphics[width=0.4\linewidth, height = 5cm]{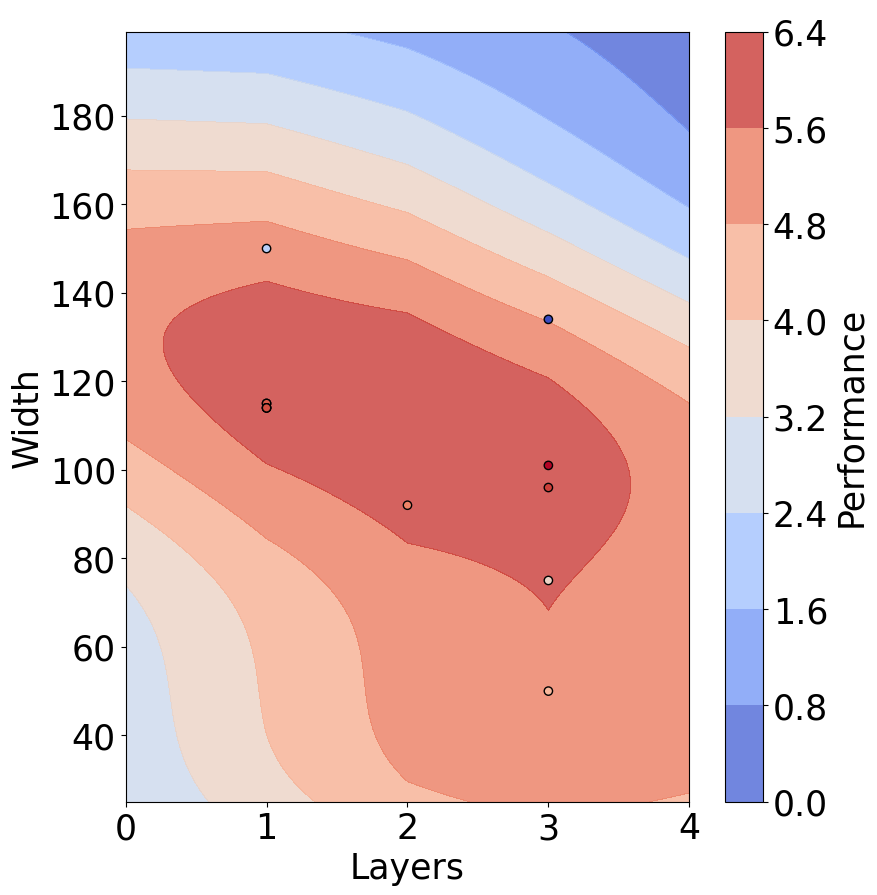}
    \caption{Posterior distribution of the network performance on the \emph{log(E/MeV)} and centroid (left) and \emph{log(E/MeV)} and dispersion (right) multi-target regression as a function of layers and width per layer, having set $\eta=10^{-4}$ and $\lambda=0$.}
    \label{fig:net_opt_1}
\end{figure}

\begin{figure}[H]
    \centering
    \includegraphics[width=0.4\linewidth, height = 5cm]{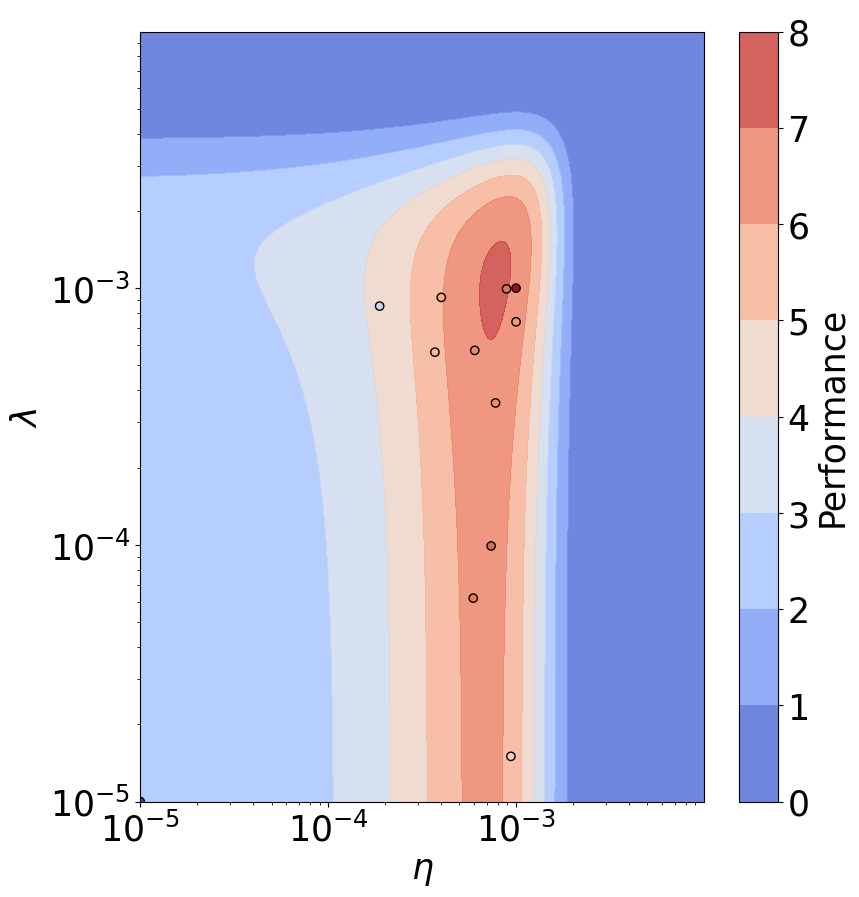}
    \hspace{1cm}
    \includegraphics[width=0.4\linewidth, height = 5cm]{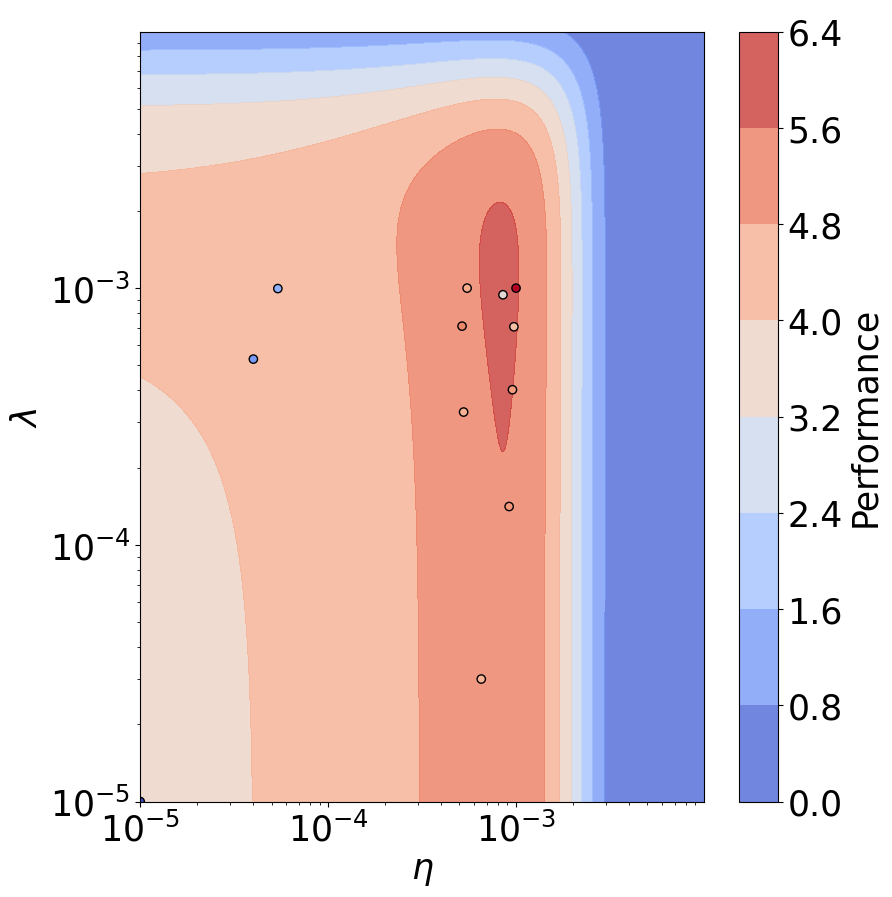}
    \caption{Posterior distribution of the network performance on the \emph{log(E/MeV)} and centroid (left) and \emph{log(E/MeV)} and dispersion (right) multi-target regression learning rate $\eta$ and weight regularization constant $\lambda$, having set the number of layers $N=2$ and the network width $W=120$, obtained from architecture optimization above.}
    \label{fig:net_opt_sigma}
\end{figure}

\section{Nanophotonic Hardware Implementation}
\label{sec:hardware_Implementation}





As described in the introduction, the speed of the processes to be sampled and analysed need sub-nanosecond resolution, putting severe constraints on the hardware solutions. As an additional parameter, energy consumption  should be kept low to not increase the overall energy budget of the experiment too much. Most neuromorphic systems operate at up to 200 MHz~\cite{javanshir2022}, which is equivalent to 5 ns cycle time, not including the light conversion step.

We propose a nanophotonic implementation for computation, with all electrical and optical components integrated in a single chip stack including optical detectors. III-V semiconductors, with their excellent photon/electron conversion efficiencies are aptly suited for this. Nanowire (NW) devices made from III-Vs can detect light, emit light, and also operate as electronic components~\cite{barrigon2019}. III-V NWs have the additional advantage of having excellent cross-sections for light absorption and due to their small size can operate at very fast timescales with very low energy consumption~\cite{winge2020, wittenbecher2021}. Combining these functions in InP NWs, neural networks using light for communication has recently been proposed~\cite{winge2020, Winge2023}, and light communication between individual NW components demonstrated~\cite{flodgren2025}.

These components and developments thereof can also be used in the present proposal, but a range of other III-V NWs with direct bandgap are also relevant~\cite{barrigon2019}. While  individual NW devices were employed for neural network implementation~\cite{flodgren2025},  arrays of parallel NWs can also be fabricated, delivering similar performance in terms of sensitivity and operational times, however with larger currents~\cite{Alcer2023}. The higher current leads to a more robust circuit while also having a larger area of detection for each pixel. A photodetector circuit that can also act as a sigmoid neuron then consists of two different types of photodiodes that supply or remove charge at the gate of an InAs NW field effect transistor (FET) which then controls an InP NW LED emitter~\cite{Winge2023}. Operated at suitable voltages, inputs on the two photodiodes summarize as exciting/inhibiting signals respectively, which is treated via a sigmoid function to result in (possible) LED light emission. Because the gate turns on when a sufficient charge has been accumulated, pulses do not need to be strictly timed, and in principle it can work in a spiking mode. Spiking NW neurons including emitters have been developed~\cite{romeira2013}, and spiking photonic neurons with picoseconds have been suggested~\cite{Shastri2015}. 

For the present detector and analysing systems using III-V NWs  appears as an interesting solution as they can act both as detectors of light and artificial neurons including LIF units. Further, using light signals for network connections is highly relevant as this will give the fastest possible connectivity. The 450-550 nm bandwidth light emission of Lead Tungstate can be detected by the numerous III-V materials with bandgaps below 2.25 eV, including GaAs, InP and various ternaries of these two compounds. 

\begin{figure}
    \centering
    \includegraphics[width=0.98\linewidth]{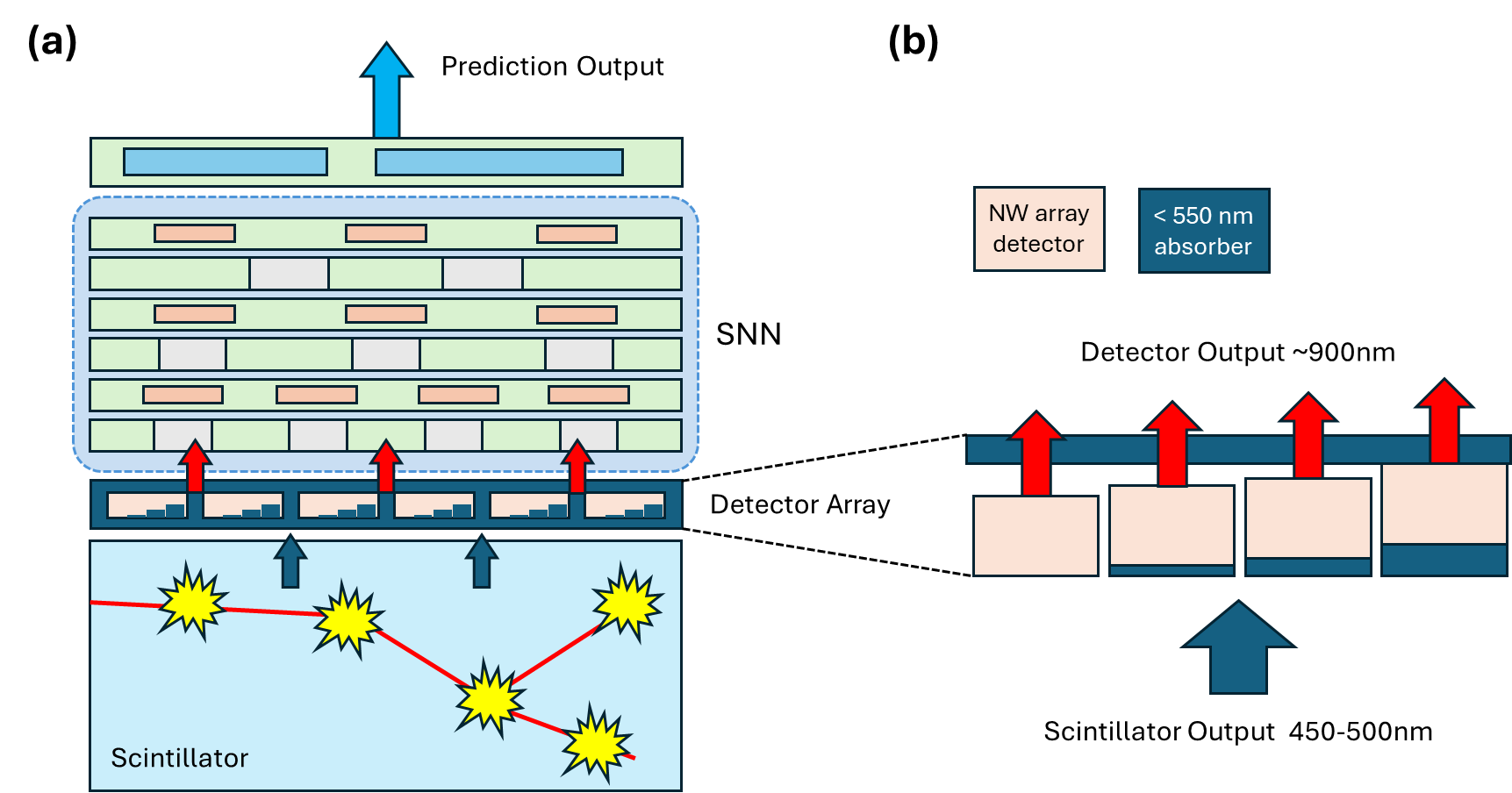}
    \caption{(a) Possible implementation of a readout system for a calorimeter block (bottom left) -- light pulses are received by the layer of detector arrays and passed onto the SNN, which comprises of multiple layers of metalens/waveguide broadcasting and LIF neurons. Finally, the prediction output from the SNN is decoded and read out. (b) 4 channel detector superpixel, with individual detectors receiving progressively attenuated signals due to their respective thicker absorbers.}
    \label{fig:hardware}
\end{figure}

A layered system of arrays of NW components organized in a suitable geometry, where the bottom layer absorbs the incident scintillator light and the top layer serves as output of the processed information, can be considered as a trainable neuromorphic system. Referring to such a system in Fig.~\ref{fig:hardware}(a), we first turn to the light detectors. Here 4 detectors in each pixel should be constructed that sequentially have an order of magnitude different sensitivity as described in Eq. \ref{eqn:spike}. A suitable NW neuron architecture has one phototransistor for excitation coupled with a NW FET and NW LED emitter~\cite{Winge2023}. When sufficient light hits the phototransistor, the LED starts to emit. A simple way to achieve different thresholds on the four different phototransistors is to place scintillator-light absorbing  GaInP layers of varying thickness between the scintillator and respective NW detectors. As 1 µm thick GaInP will be enough to completely absorb the scintillator light output and absorption decreases exponentially, varying thickness can be directly used to get the required orders of magnitude difference. A representative design is shown in Fig.~\ref{fig:hardware}(b). Operating the NW LED emitter at lower energies (such as by using InP) can ensure that the signal output from the detector neuron will not be absorbed by the GaInP and reach the SNN. The operating timescales will be determined by the sensing and emission components of the neuron. The signal transmission is at the speed of light and can be assumed to be instantaneous at these length scales (light travels 1 µm in $\sim3$ fs). Although the phototransistor risetime is $\sim10$ ps, the longer NW LED risetime of $\sim100$ ps sets the speed of the system. This response time is sufficient for the time window required in the present study.

The light signals from the four detector superpixel can be guided further upward towards the SNN layer using waveguiding structures of dielectric or plasmonic nature. For the next layer, an integrated LIF NW array neuron can be placed to act as the first layer in the SNN. Here either a rate dependent or spiking schemes can be used as has been developed using various III-V materials~\cite{romeira2013, Winge2023}. To create the weight between different layers of NW neurons, light signals should be manipulated to achieve varying intensity in different directions, using several wavelengths or polarization. NWs are sensitive to all of these depending on materials and geometric structure with sub-wavelength sensitivity possible~\cite{barrigon2019, marsell2018}. Metalenses can also be used for inter-connectivity and to assign weights~\cite{Meng2021}. One can work with pre-training during the design phase and use fixed weights. Alternatively, absorbing molecules can be introduced that can be used as artificial synapses in the connections between the NWs~\cite{Alcer2025}.

\section {Conclusions}
\label{sec:conclusion}

While a high-granularity of calorimeters appears today as a dire necessity for cutting-edge future HEP applications and studies at the high-energy and high-intensity frontiers, the fast readout and creation of suitable trigger primitives, as well as the size of the generated data volume and energy consumptions, pose problems whose solution lays beyond the state of the art of the relative technologies. Processing the vast amounts of spatiotemporal data generated by light emission within sub-nanosecond time frames is challenging. Traditional machine learning approaches suffer from high latency and energy dissipation, while FPGA-based solutions are often limited by fixed logic constraints.

In this article we propose a way forward by using neuromorphic computing for the extraction of fast topological primitives from the pattern of energy releases in a scintillating medium. The absence of incoming light amplification, and the intrinsic capability of the considered system to work with photonic information processing, make it a promising solution to the aforementioned bottlenecks. 

Our studies indicate that an array of neurons exchanging spikes generated by photon intensities and arrival times on an array of input sensors can extract useful information on the total energy release, its centroid, and its dispersion, opening the way to the creation of high-level features that may extract from the calorimeter block the same amount of information that would be harvested by a highly granular instrument, thus bypassing the related readout and technology challenges.

The neuromorphic approach, particularly when implemented in nanophotonic hardware, is well suited for this problem. By leveraging event-driven computation and sub-nanosecond neurosynaptic dynamics, neuromorphic systems can efficiently extract relevant features from the evolving light patterns in a homogeneous calorimeter without requiring high segmentation. The proposed III-V NW-based implementation further enhances performance by enabling ultra-fast and energy-efficient spike-based processing, paving the way for real-time, low-power event reconstruction in high-energy physics detectors.

\funding{
The work by TD and FS was partially supported by the Wallenberg AI, Autonomous Systems and Software 
Program (WASP) funded by the Knut and Alice Wallenberg Foundation. 
The work by MA and FS was partially supported by the Jubilee Fund at the Lule{\aa} University of Technology.
The work by PV was supported by the ``Ramón y Cajal” program under Project No. RYC2021-033305-I funded by the MCIN MCIN/AEI/10.13039/501100011033 and by the European Union NextGenerationEU/PRTR.
JK is supported by the Alexander-von-Humboldt foundation.
The work by AD and AM was supported by European Union Horizon Europe project InsectNeuroNano (Grant 101046790) and the Wallenberg Initiative Material Science for Sustainability (WISE) funded by the Knut and Alice Wallenberg Foundation. 
}

\begin{adjustwidth}{-\extralength}{0cm}

\reftitle{References}
\bibliography{main.bib}

\PublishersNote{}
\end{adjustwidth}
\end{document}